\documentclass{moriond}

\def\be{\begin{equation}}
\def\ee{\end{equation}}
\def\bea{\begin{eqnarray}}
\def\eea{\end{eqnarray}}

\usepackage{lineno}
\usepackage{graphicx}
\usepackage{calc}

\usepackage[figuresright]{rotating}

\newcommand{\pT}{\ensuremath{p_{\rm T}}}
\newcommand{\MeVc}{\ensuremath{{\rm MeV/}c}}
\newcommand{\GeVc}{\ensuremath{{\rm GeV/}c}}

\newcommand{\PbPb}{Pb--Pb}

\newcommand{\sqrtsNN}{\ensuremath{\sqrt{s_{\rm NN}}}}

\newcommand{\cTeV}{5.02~TeV}

\newcommand{\sqrtsNNcTeV}{\sqrtsNN~=~\cTeV}

\newlength{\lengtha}
\newlength{\lengthb}
\title{Measuring hydrodynamical expansion via the production of identified hadrons in \PbPb\ collisions with ALICE}
\author{Nicol\`o Jacazio (Bologna University and INFN) for the ALICE Collaboration}

\begin{document}
  \maketitle
  \begin{abstract}
  During the LHC Run-2, ALICE has collected data from Pb--Pb collisions at $\sqrt{s_{\rm{NN}}}$~=~5.02 TeV. 
  The centrality dependence of identified particle production, including elliptic ($v_{2}$) and higher harmonic flow coefficients ($v_{3},v_{4}$), has been measured.
  The high-precision measurement of transverse momentum (\pT) differential elliptic flow of the $\phi$-meson (whose mass is close to that of the proton) allows for a unique testing of mass ordering at low \pT\ as well as baryon and meson grouping at intermediate \pT . 
  The \pT-differential hadron spectra are presented and, together with flow coefficients, compared with state-of-the-art calculations from models based on relativistic hydrodynamics coupled with UrQMD.
  The added transport code is to describe rescattering in the hadronic phase, which has been successful in describing the \pT -spectra of identified particles up to a few \GeVc . 
  Moreover, results from the simultaneous Blast-Wave fit to the \pT\ distributions are compared across multiple collisions energies and system sizes in order to address the evolution of collective behaviour from small systems to large systems.
\end{abstract}

  \section{Introduction}
  \label{sec:Introduction}
  The ultimate goal of heavy-ion physics is the study of the properties of the Quark-Gluon Plasma (QGP), a de-confined state of matter in which chiral symmetry is restored.
  The presence of a de-confined phase manifests with typical signatures that can be quantified by studying the particle production.
  In particular the measurement of identified particles provides a unique way to gain insight into the physical quantities at play.
  The latest \PbPb\ runs at the LHC, concluded in 2015, recorded \PbPb\ collisions at the highest energy ever achieved in the laboratory, allowing the quantitative comparison with collisions at lower energy.
  The ALICE experiment~\cite{ALICEExp,ALICEperf}, thanks to its excellent tracking performance coupled with extensive particle identification (PID) capabilities, is particularly well suited for the study of identified hadron production over a wide range of transverse momentum. 
  This is achieved by combining multiple techniques that allow, at mid-rapidity, to perform hadron identification starting from 100 \MeVc\ and up to $\sim 20$ \GeVc .
  \section{Data analysis and results}
  \label{sec:Results}
  We report results on the production of identified $\pi$, K, p and $\phi$-meson production measured in \PbPb\ collisions at \sqrtsNNcTeV\ as a function of centrality.
  The data sample was recorded in 2015 with a minimum-bias trigger.
  The total charge collected in the V0 detectors (V0M amplitude), a set of two scintillator hodoscopes located in the pseudorapidity region $2.8 < \eta < 5.1$ (V0A) and $- 3.7 < \eta < - 1.7$ (V0C) and covering the full azimuth, was used to determine the centrality of each \PbPb\ collision defined as percentiles of the total hadronic cross section.
  Further details on the centrality determination can be found in~\cite{ALICECent5TeVEstimation,ALICECent276TeVEstimation}.
  The amount of pile-up per event was reduced by selecting runs with low interaction rate and by rejecting events with more than one reconstructed vertex, resulting in a negligible contamination effect.
  \begin{figure}
  \centering
  \begin{minipage}{\textwidth}
    \centering
    \begin{minipage}{.34\textwidth}
      \centering
      \includegraphics[trim = 0 0 20 0, clip, width=\textwidth]{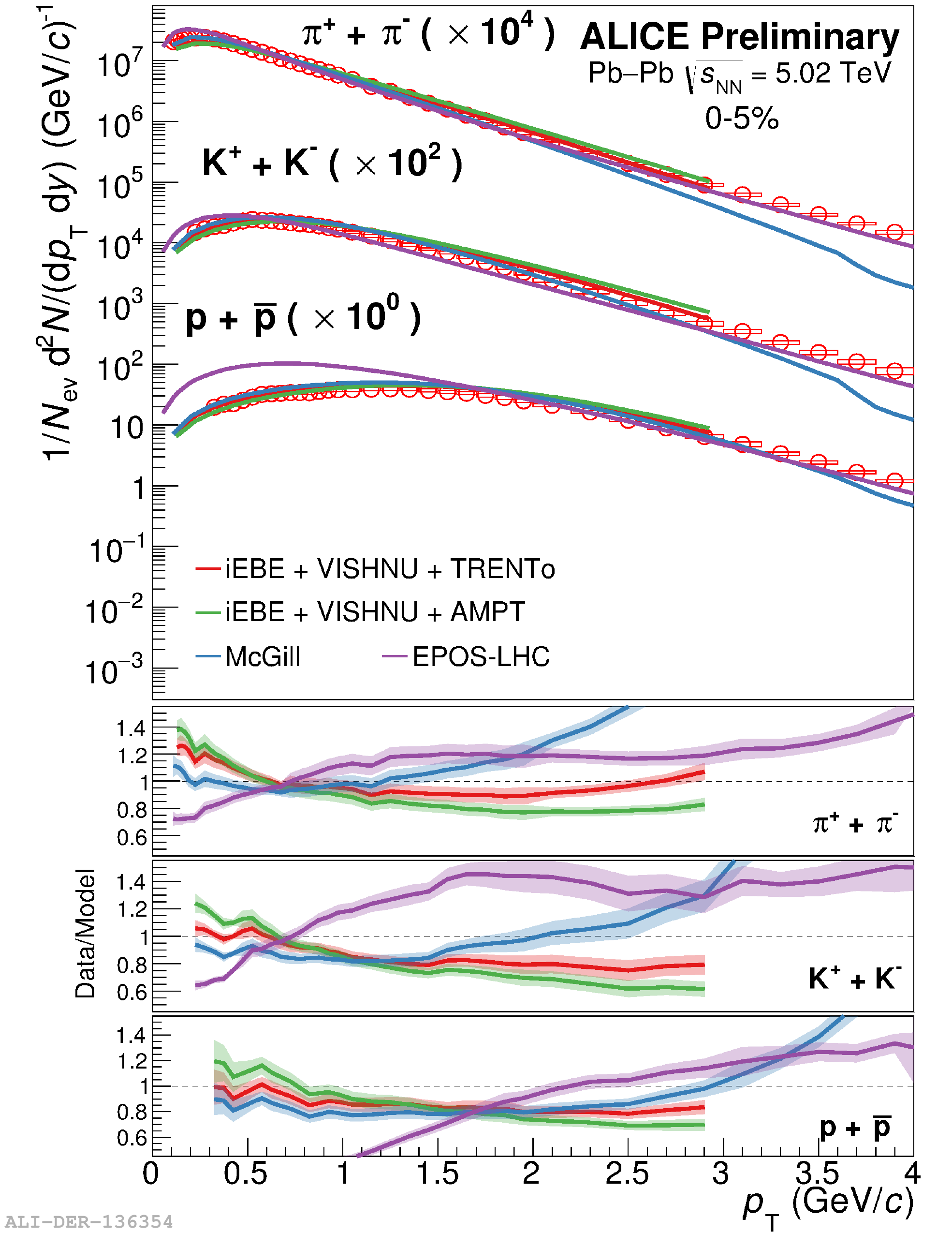}\\
      (a)
      \vskip .1cm
      \includegraphics[trim = 0 0 20 0, clip, width=\textwidth]{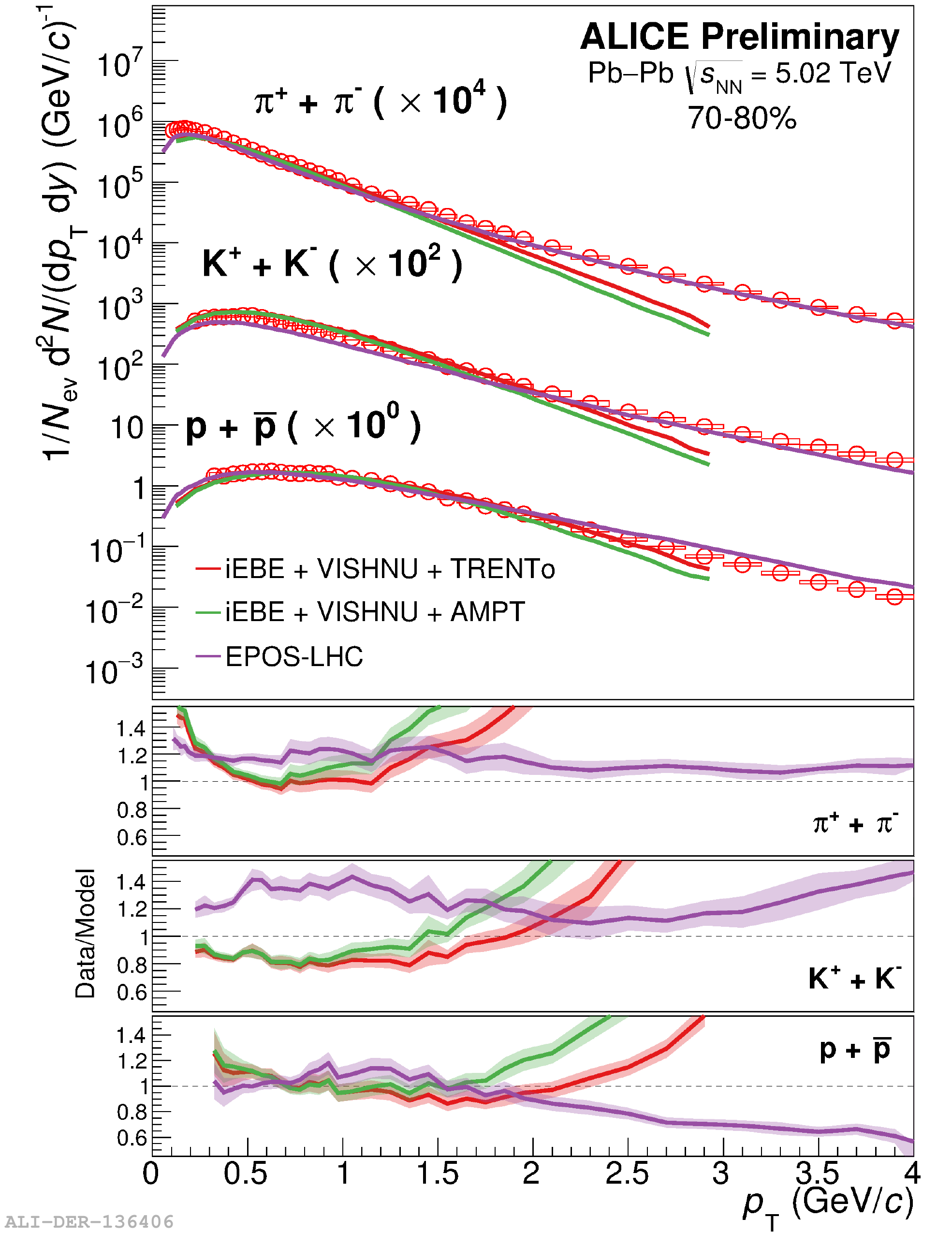}
      (c)
    \end{minipage}%
    \hspace{.01\textwidth}
    \begin{minipage}{.62\textwidth}
      \centering
      \vspace*{-.4cm}
      \includegraphics[trim = 3 3 16 36, clip, width=\textwidth]{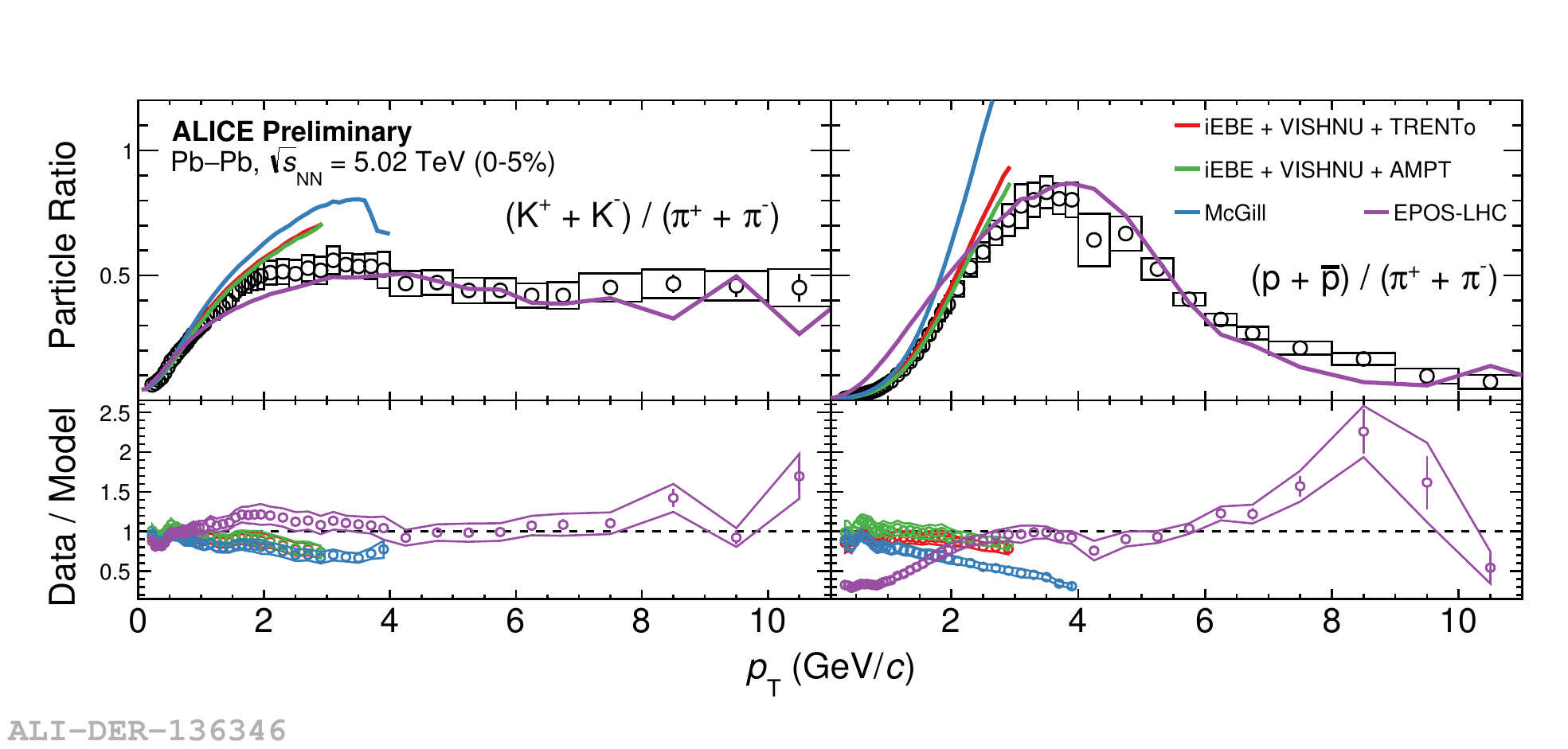}\\
      \vspace*{-.4cm}
      (b)\\
      \vspace*{.1cm}
      \includegraphics[trim = 3 3 16 36, clip, width=\textwidth]{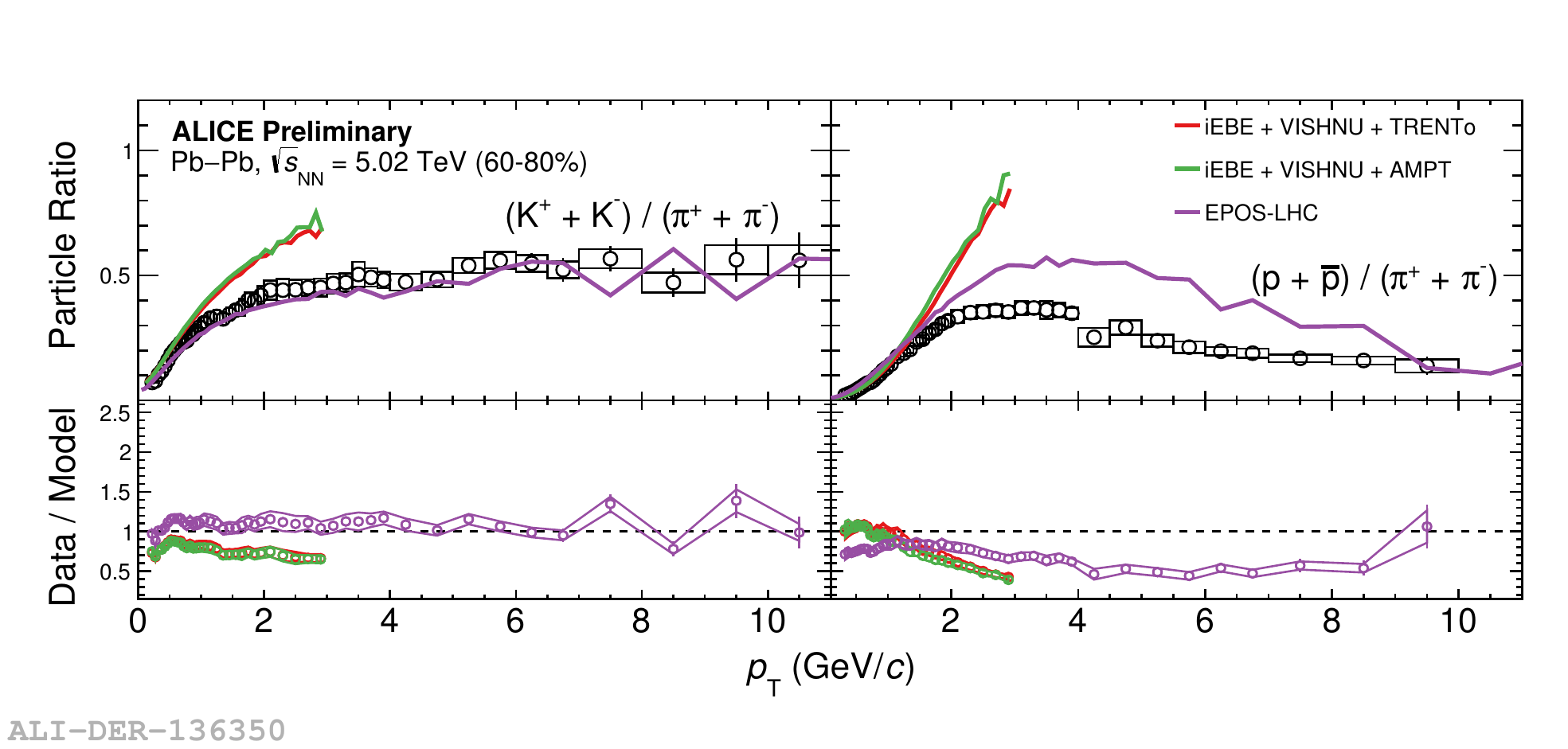}\\
      \vspace*{-.4cm}
      (d)\\
      \vspace*{.1cm}
      \includegraphics[trim = 3 3 50 12, clip, width=.85\textwidth]{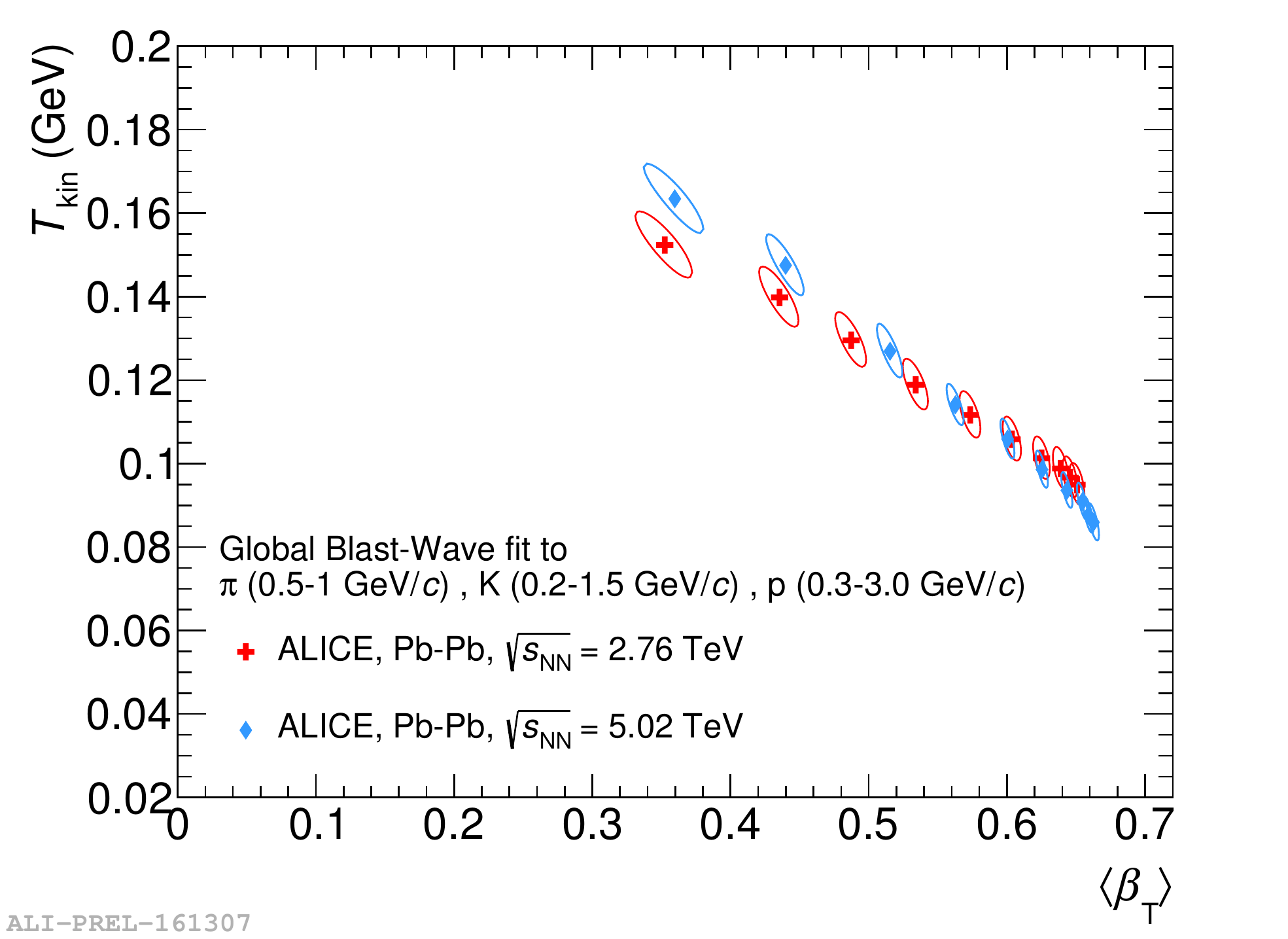}\\
      \vspace*{-.4cm}
      (e)
    \end{minipage}
    \hspace{.01\textwidth}
    
    \hspace{.01\textwidth}
    
    \caption{
      \label{fig:IDspectra}
      Spectra of identified charged pions, kaons and protons for $0-5\%$ (a) and $70-80\%$ (c) as a function of \pT\ as measured in \PbPb\ collisions at \sqrtsNNcTeV\ and compared to model predictions (iEBE + VISHNU, McGill, EPOS-LHC).
      (b) and (d) show particle ratios compared to the predictions from the same models.
      In figure (e) are reported the values of the Blast-Wave parameters obtained by performing a combined fit to the $\pi$, K, p spectra for each centrality class across different collision energy. 
    }
  \end{minipage}
\end{figure}

  The \pT\ spectra of identified $\pi$, K and p are shown for two centrality classes in Figs. \ref{fig:IDspectra}(a) and \ref{fig:IDspectra}(c).
  This measurement refers to primary particles~\cite{ALICEPrim} only, i.e. the contributions from weak decays of strange particles and from particle knock-out in the material were removed with the data driven approach described in~\cite{ALICEcentral276}.
  The systematic uncertainties were estimated by varying the PID techniques and the selection criteria used to define the track sample. 
  The evaluation of the efficiencies and acceptance corrections was performed by using events simulated with the HIJING~\cite{HIJING} event generator and embedded into a detailed description of the ALICE detector through which tracks are propagated with the GEANT3~\cite{GEANT3} transport code.
  A direct comparison of the spectral shapes between the two centralities reveals steeper falling spectra for the peripheral case, while for the central one particles are boosted towards higher \pT\ following the mass hierarchy typical of radial flow.
  The spectral shapes are compared to the predictions from different models such as iEBE-VISHNU with TRENTo and AMPT initial conditions~\cite{Vishnu,AMPT}, IP-Glasma + Music + UrQMD (McGill)~\cite{McGill} and EPOS-LHC~\cite{EPOSlhc}.
  In central collisions, the description of the data is qualitatively good especially at low \pT\ where hydrodynamics is expected to work the best.
  EPOS-LHC is found to give a better description of the peripheral collisions.
  The particle ratios p/$\pi$ and K/$\pi$ are computed and compared to the predictions of the same models in Figs. \ref{fig:IDspectra}(b) and \ref{fig:IDspectra}(d).
  In central collisions the agreement is remarkably good at low \pT\ (below 2 \GeVc ) for the K/$\pi$ ratio (meson/meson), the ratio is also well reproduced by EPOS-LHC up to high \pT; on the contrary in peripheral collisions only EPOS-LHC is able to reproduce the data well.
  The p/$\pi$ ratio (baryon/meson) is well described at low \pT\ by the iEBE-VISHNU model, while McGill shows increasing discrepancies and EPOS-LHC agreement is good only above 2 \GeVc . 
  In peripheral collisions the iEBE-VISHNU model describes the data below 1 \GeVc\ and EPOS-LHC generally overestimates the value of the p/$\pi$ ratio.
  The picture depicted by the models indicates that hydrodynamics is at work in central heavy ion collisions and is able to describe the data, while in peripheral collisions the description is more difficult and a core-corona such as EPOS-LHC is able to better describe the data.
  These observations advocate the creation of a fireball subject to larger flow in central heavy-ion collisions while in peripheral collisions hydrodynamical models have more difficulty to describe the data.
  Similar conclusions can be drawn from the analysis of the spectral shape in the framework of the Blast-Wave model~\cite{bgbw}, shown in Fig. \ref{fig:IDspectra}(e), which yields a larger transverse expansion velocity for the most central events.
  \begin{figure}
  \centering
  \includegraphics[trim = 0 0 18 0, clip, width=\textwidth]{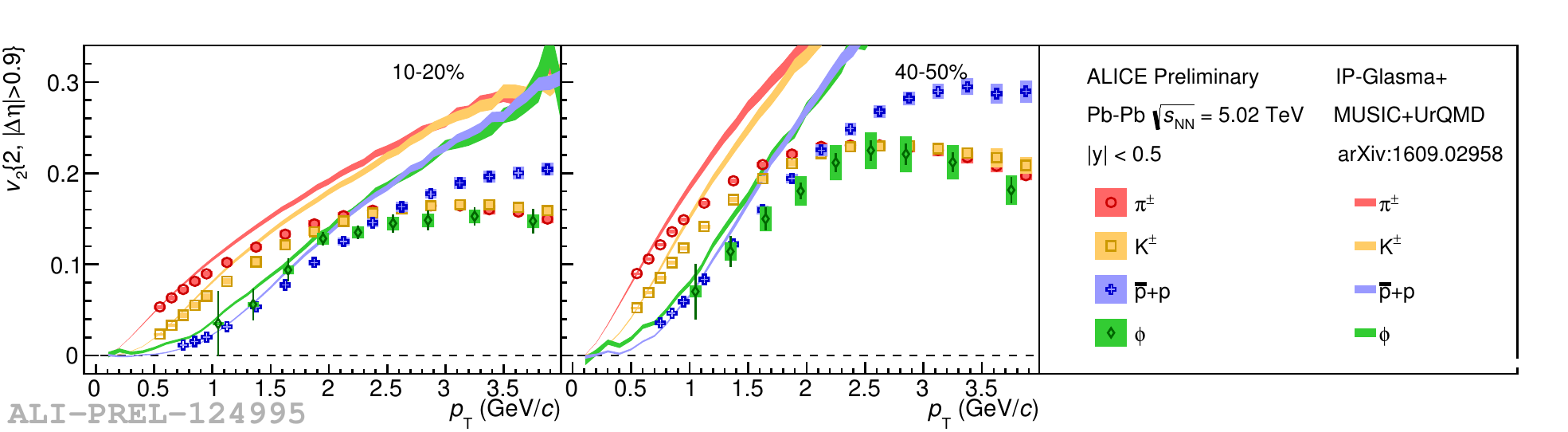}\\
  (a)\\
  \includegraphics[trim = 0 0 18 0, clip, width=\textwidth]{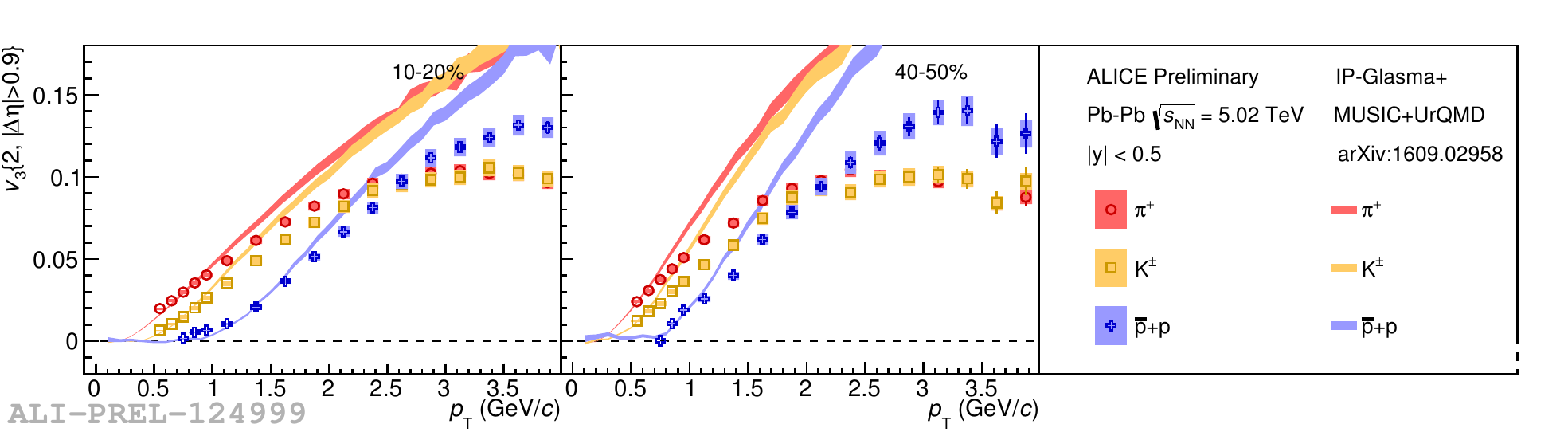}\\
  (b)\\  
  \caption{ \label{fig:vNhydro23} 
    $v_{2}$ (a) and $v_{3}$ (b) of identified pions, kaons, protons, and $\phi$-mesons for semi-central ($10-20\%$) and semi-peripheral ($40-50\%$) centrality classes, measured with the scalar product method.
    The measured flow coefficients are compared with the  IP-Glasma + Music + UrQMD (McGill) model predictions~$^{11}$.
  }
\end{figure}
\begin{figure}
  \centering
  \includegraphics[trim = 0 0 18 0, clip, width=\textwidth]{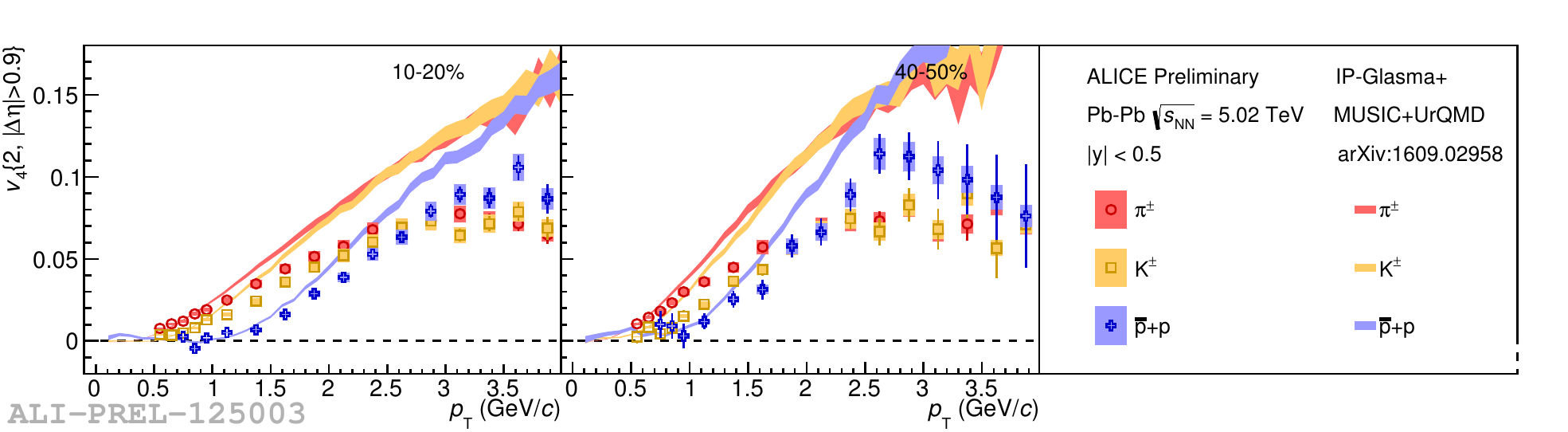}\\
  \caption{ \label{fig:vNhydro4} 
    $v_{4}$ of identified pions, kaons, protons for semi-central ($10-20\%$) and semi-peripheral ($40-50\%$) centrality classes, measured with the scalar product method.
    The measured flow coefficients are compared with the  IP-Glasma + Music + UrQMD (McGill) model predictions~$^{11}$.
  }
\end{figure}

  In a complementary way, particle production can be studied more differentially by measuring the flow coefficients obtained studying the particle azimuthal distribution with respect to the common symmetry plane.
  In Figs. \ref{fig:vNhydro23} and \ref{fig:vNhydro4}, the results of the flow coefficients $v_2$, $v_3$ and $v_4$ as a function of \pT\ as measured with the scalar product method in Pb--Pb collisions at \sqrtsNNcTeV\ are shown.
  More details can be found in~\cite{ALICEflow}.
  The conclusions drawn from Figs. \ref{fig:vNhydro23} and \ref{fig:vNhydro4} complete the ones on the spectral shape.
  At low \pT , $v_2$ shows a clear mass ordering which can be once more explained in terms of strong radial flow, for $\pT\ > 3$ \GeVc\ the $v_2$ shows a separation of baryons from mesons suggesting that the production of hadrons may happen via quark coalescence.
  The McGill~\cite{McGill} IP-Glasma model prediction (viscous hydrodynamics with $\eta/s = 0.095$ and temperature-dependent $\zeta/s(T)$ coupled to a hadronic cascade) of the $v_n$ are also shown in Figs. \ref{fig:vNhydro23} and \ref{fig:vNhydro4}.
  As for the comparison to the measured particle spectra the model seems to describe flow coefficients better in central collisions for $\pT\ < 1$, while an overestimation is observed in the same \pT\ range in more peripheral ones.
  When performing this comparison it is important to note that in the theoretical models the breaking of the mass ordering for the $\phi$-meson occurs during the hadronic re-scattering phase.
  \section{Conclusions}
  \label{sec:conclusions}
  The ALICE Collaboration has presented the results on the production of identified pions, kaons, protons and $\phi$-mesons measured as a function of the event centrality in \PbPb\ collisions at \sqrtsNNcTeV .
  Particle flow was quantified thanks to the analysis of the spectral shape in the Blast-Wave framework and the direct comparison to models as well as the measurement of $v_2$, $v_3$ and $v_4$ coefficients and their comparison to the model predictions.
 
  \section*{References}
  \bibliography{bibliography}
  
\end{document}